\documentclass[twocolumn,english,aps,prd,nofootinbib]{revtex4-2}
\usepackage{lmodern}
\usepackage[LGR,T1]{fontenc}
\usepackage[latin9]{inputenc}
\usepackage{xcolor}
\usepackage{babel}
\usepackage{amsmath}
\usepackage{amsthm}
\usepackage{amssymb}
\usepackage{graphicx}
\usepackage[unicode=true]
 {hyperref}

\makeatletter

\DeclareRobustCommand{\greektext}{%
  \fontencoding{LGR}\selectfont\def\encodingdefault{LGR}}
\DeclareRobustCommand{\textgreek}[1]{\leavevmode{\greektext #1}}
\ProvideTextCommand{\~}{LGR}[1]{\char126#1}

\theoremstyle{plain}
\newtheorem{thm}{\protect\theoremname}
\theoremstyle{remark}
\newtheorem{rem}[thm]{\protect\remarkname}

\usepackage{enumerate}

\makeatother

\providecommand{\remarkname}{Remark}
\providecommand{\theoremname}{Theorem}

\begin{document}
\title{Impact of Star Pressure on $\gamma$ in Modified Gravity beyond Post-Newtonian
Approach}
\author{Hoang Ky Nguyen$\,$}
\email[\ ]{hoang.nguyen@ubbcluj.ro}

\affiliation{Department of Physics, Babe\c{s}-Bolyai University, Cluj-Napoca 400084,
Romania}
\author{Bertrand Chauvineau$\,$}
\email[\ ]{bertrand.chauvineau@oca.eu}

\affiliation{Universit\'e C\^ote d'Azur, Observatoire de la C\^ote d\textquoteright Azur,
CNRS, Laboratoire Lagrange, Nice cedex 4, France}
\date{July 1, 2024}
\begin{abstract}
\vskip2pt We offer a concrete example exhibiting marked departure
from the Parametrized Post-Newtonian (PPN) approximation in a modified
theory of gravity. Specifically, we derive the exact formula for the
Robertson parameter $\gamma$ in Brans-Dicke gravity for spherical
compact stars, explicitly incorporating the pressure content of the
stars. We achieve this by exploiting the \emph{integrability} of the
$00-$component of the Brans-Dicke field equation. In place of the
conventional PPN result $\gamma_{\,\text{PPN}}=\frac{\omega+1}{\omega+2}$,
we obtain the analytical expression $\gamma_{\,\text{exact}}=\frac{\omega+1+(\omega+2)\Theta}{\omega+2+(\omega+1)\Theta}$
where $\Theta$ is the ratio of the total pressure $P_{\parallel}^{*}+2P_{\perp}^{*}$
and total energy $E^{*}$ contained within the star. The dimensionless
quantity $\Theta$ participates in $\gamma$ due to the scalar degree
of freedom of Brans-Dicke gravity. Our \emph{non-perturbative} formula
is valid for all field strengths and types of matter comprising the
star. In addition, we establish two new mathematical identities linking
the active gravitational mass, the ADM (Arnowitt-Deser-Misner) mass,
and the Tolman mass, applicable for Brans-Dicke gravity. We draw four
key conclusions:\linebreak (1) The usual $\gamma_{\,\text{PPN}}$
formula is violated for high-pressure mass sources, such as neutron
stars, viz. when $\Theta\neq0$, revealing a limitation of the PPN
approximation in Brans-Dicke gravity. (2) The PPN result mainly stems
from the assumption of pressureless matter. Even in the weak-field
star case, non-zero pressure leads to a violation of the PPN formula
for $\gamma$. Conversely, the PPN result is a good approximation
for low-pressure matter, i.e. when $\Theta\approx0$, for all field
strengths. (3) Observational constraints on $\gamma$ set \emph{joint}
bounds on $\omega$ and $\Theta$, with the latter representing a
global characteristic of a mass source. If the equation of state of
matter comprising the mass source approaches the ultra-relativistic
form, entailing $\Theta\simeq1$, $\gamma_{\,\text{exact}}$ converges
to $1$ \emph{irrespective} of $\omega$. More generally,
regardless of $\omega$, ultra-relativistic matter suppresses the
scalar degree of freedom in the exterior vacuum of Brans-Dicke stars,
reducing the vacuum to the Schwarzschild solution. (4) In a broader
context, by exposing a limitation of the PPN approximation in Brans-Dicke
gravity, our findings indicate the significance of considering the
interior structure of stars in observational astronomy when testing
candidate theories of gravitation that involve additional degrees
of freedom besides the metric tensor.
\end{abstract}
\maketitle

\section{\label{sec:Motivation}Motivation}

The parametrized post-Newtonian (PPN) framework has been an invaluable
tool in the study of gravitational theories \citep{Will-2011,Will1,Will2}.
It is founded upon the assumptions of weak field and slow motion,
representing the post-Newtonian limit of gravity. In this limit, the
framework characterizes a given metric theory, at the first post-Newtonian
level, by a set of ten real-valued parameters.\vskip4pt

One of the most powerful applications of the PPN formalism is the
calculation of the Robertson (or Eddington-Robertson-Schiff) parameter
$\gamma$. This important parameter governs the amount of space-curvature
produced by a body at rest and can be directly measured via the detection
of light deflection and the Shapiro time delay. For one of the simplest
extensions beyond General Relativity (GR)---the Brans-Dicke (BD)
theory---the PPN approach is known to yield
\begin{equation}
\gamma_{\,\text{PPN}}=\frac{\omega+1}{\omega+2}\label{eq:PPN-gamma}
\end{equation}
where $\omega$ is the BD parameter. This formula recovers the result
$\gamma_{\,\text{GR}}=1$ known for GR in the limit of infinite $\omega$,
in which the BD scalar field approaches a constant field. Current
bounds using Solar System observations set the magnitude of $\omega$
to exceed $40,000$. Generalizations of the PPN $\gamma$ result to
other modified theories of gravity are available \citep{Hohmann-2014-b,Hohmann-2014-a,Hohmann-2015,Hohmann-2016,Hohmann-2017-a,Hohmann-2017-b,Hohmann-2019,Hohmann-2020-a,Hohmann-2020-b,Hohmann-2020-c,Hohmann-2020-d,Hohmann-2021,Hohmann-2022,Hohmann-2023,Will1,Will2}.\vskip4pt

In the case of GR, by virtue of Birkhoff's theorem, spherically symmetric
vacuum solutions are static (discarding the black hole region in the
case of a fully vacuum spacetime) and asymptotically flat. The vacuum
spacetime exterior of a mass source is described by the Schwarzschild
metric which is dependent on only one parameter, the Schwarzschild
radius. All information regarding the internal structure and composition
of the source, namely, the types of matter comprising it as well as
the distribution profile of matter in the source, is fully encapsulated
in the Schwarzschild radius. Since GR, as a \emph{classical} theory,
lacks an inherent length scale (such as the Planck length), the dimensionful
Schwarzschild radius cannot participate in the \emph{dimensionless}
$\gamma$ parameter. Thus, this parameter is independent of the source
in GR (a fact compatible with $\gamma_{\,\text{GR}}=1$).\vskip4pt

BD gravity, however, has a richer structure due to the BD scalar field
$\Phi$, an additional degree of freedom besides the metric components.
For the special case of black holes, the no-hair theorem applies \citep{Hawking-1972-BD,SotiriouFaraoni-2012},
meaning that, in the non-rotating case, the vacuum exterior to a static
BD black hole is the Schwarzschild solution. Consequently, black holes
in BD gravity, \emph{irrespective of the value of $\omega$}, have
$\gamma=1$ rather than the usual $\frac{\omega+1}{\omega+2}$ result.
Besides black holes, BD gravity exhibits other structures---normal
stars and exotic ones, such as wormholes and naked singularities \citep{Nguyen-2023-WEC,Agnese-1995,Faraoni-2016}.
For these structures, the BD scalar field in the exterior vacuum is
generally non-constant. The PPN formalism makes use of the \emph{non-constancy}
in $\Phi$ to derive the usual PPN result for stars, given in Eq.
\eqref{eq:PPN-gamma}, which explicitly depends on the BD parameter
$\omega$ \citep{Weinberg}. Yet, it is important to note that the
post-Newtonian (PN) $\gamma$ parameter, per Eq. \eqref{eq:PPN-gamma},
contains no information about the star (aside from the fact that the
star is regular at its center), as the BD parameter $\omega$ is a
parameter of the theory, not one of the resulting exterior vacuum.\vskip4pt

In contrast to the second-derivative GR, where all information about
the stellar source is condensed into one single parameter---the Schwarzschild
radius---BD gravity, as a higher-order theory, permits the internal
structure of the star to influence higher-derivative characteristics
of the exterior solution, potentially leaving its footprints on the
PN parameters. In their seminal work \citep{BransDicke-1961}, Brans
and Dicke recognized the approximate nature of estimating the two
parameters characterizing the exterior vacuum solution, later known
as the Brans Class I solution \citep{Brans-1962}. This recognition
was evidenced in Eq. (34) of Ref. \citep{BransDicke-1961}. The PPN
formula in Eq. \eqref{eq:PPN-gamma} should be regarded as an approximation
applicable in the limit of weak field everywhere (including inside
the star) and slow motion in BD gravity, whereby the higher-derivative
features of the exterior vacuum are suppressed. An important question
arises: \emph{Can the internal structure of a star in BD gravity,
and in modified theories in general, manifest through the PN parameters?
And if so, under what conditions?}\vskip4pt

An obvious course of action would be to lift the weak field hypothesis
on the source of the field, enabling an exact calculation of the external
field, from which the PN parameters can be extracted. An attempt in
this direction has been made recently in \citep{Faraoni-2020} where
it is suggested that the higher-order terms can impact the $\gamma$
parameter. Yet, there exists another condition, albeit less explicit,
related to the pressure within the star. Note that the PPN requirement
of slow motion applies not only for macroscopic objects but also for
their microscopic constituents. Per the post-Newtonian bookkeeping
scheme in Ref. \citep{Will1}, this translates to a requirement for
low pressure relative to the energy content. Whereas a star is stationary,
\emph{the microscopic motion of the matter contained within its domain
can be relativistic}, resulting in appreciably high pressure compared
with its energy content.\vskip4pt

Despite the challenges of relaxing the weak-field and low-pressure
constraints, significant progress can be achieved in one particular
situation---the BD theory. In our current study, we focus on a specific
case---the BD exterior vacuum surrounding a matter spherical distribution
which has a finite domain of support; we shall collectively call this
type of structure a compact star. Here, we rigorously account for
the influence of the compact mass source on the exterior vacuum and
the $\gamma$ parameter \emph{without resorting to any approximations}.
We achieve this by making use of the integrability of the 00--component
of the BD field equations, enabling us to circumvent the limitations
of the weak-field and low-pressure approximations. Advancements in
detection methods allow for the study of neutron stars, making our
investigation relevant both for practical applications and theoretical
inquiries into the formalism and methodologies employed. Our study
shall shed light on the role of stellar pressure and provide a benchmark
for assessing its impacts in theories of modified gravity.\vskip4pt

This paper provides a comprehensive account of the technical aspects
of our findings, and is structured as follows. Section \ref{sec:EMT}
revisits the form of the energy-momentum tensor (EMT) of static spherisymmetric
star sources in BD gravity. Sections \ref{sec:Standard-coord} and
\ref{sec:Isotropic-coord} handle the field equations in the standard
coordinates and transform the results to the isotropic coordinates.
The uniqueness of Brans solutions is addressed in Section \ref{sec:Unique}.
Sections \ref{sec:Matching} and \ref{sec:Robertson-params} conduct
the interior-exterior matching and derive the $\gamma$ parameter.
Section \ref{sec:Tolman-mass} obtains mass relations, valid for BD
gravity. Section \ref{sec:Higher-PPN} illustrates the fact that any
PN parameter can be calculated, by providing the explicit expression
of the 2PN parameter $\delta$. Sections \ref{sec:Discussions} and
\ref{sec:Closing} offer discussions and outlooks. A further exposition
on the uniqueness of Brans solutions is given in Appendix$\ $\ref{sec:Mapping-Brans-solutions}.

\section{\label{sec:EMT}The energy-momentum tensor}

Consider the BD action in the Jordan frame \citep{BransDicke-1961,Brans-1962}
\begin{equation}
\int d^{4}x\frac{\sqrt{-g}}{16\pi}\left[\Phi\,{\cal R}-\frac{\omega}{\Phi}\nabla^{\mu}\Phi\nabla_{\mu}\Phi\right]+\int d^{4}x\sqrt{-g}L^{(m)}
\end{equation}
with the metric signature convention $(-+++)$. Hereafter, we choose
the units $G=c=1$ with $G$ being the far-field Newtonian constant.\vskip4pt

It is well documented \citep{Bronnikov-1973} that upon the Weyl mapping
$\bigl\{\tilde{g}_{\mu\nu}:=\Phi\,g_{\mu\nu}$, $\tilde{\Phi}:=\ln\Phi\bigr\}$,
the gravitational sector of the BD action can be brought to the Einstein
frame as $\int d^{4}x\frac{\sqrt{-\tilde{g}}}{16\pi}\Bigl[\tilde{\mathcal{R}}-\left(\omega+3/2\right)\tilde{\nabla}^{\mu}\tilde{\Phi}\tilde{\nabla}_{\mu}\tilde{\Phi}\Bigr]$.
The Einstein-frame BD scalar field $\tilde{\Phi}$ has a kinetic term
with a signum determined by $(\omega+3/2)$. Unless stated otherwise,
we shall restrict our consideration to the normal (``non-phantom'')
case of $\omega>-3/2$, where the kinetic energy for $\tilde{\Phi}$
is positive.\vskip8pt

In this paper, we work exclusively in the Jordan frame. For convenience,
let us denote a rank-two tensor
\begin{equation}
X_{\mu\nu}:=\Phi\mathcal{R}_{\mu\nu}-\nabla_{\mu}\nabla_{\nu}\Phi-\frac{\omega}{\Phi}\nabla_{\mu}\Phi\nabla_{\nu}\Phi
\end{equation}
The BD field equation for the metric components is
\begin{align}
X_{\mu\nu} & =8\pi\Bigl[T_{\mu\nu}-\frac{\omega+1}{2\omega+3}g_{\mu\nu}T\Bigr]\label{eq:BD-eq-R}
\end{align}
and the equation for the BD scalar field is
\begin{align}
\square\,\Phi & =\frac{8\pi}{2\omega+3}T\label{eq:BD-eq-phi}
\end{align}
where the energy-momentum tensor (EMT) of the matter sector is
\begin{equation}
T_{\mu\nu}:=-\frac{2}{\sqrt{-g}}\frac{\delta\left(\sqrt{-g}L^{(m)}\right)}{\delta g^{\mu\nu}}
\end{equation}
For a coordinate system that is static and spherically symmetric,
the metric can be written as (with $g_{01}=g_{10}=0$)
\begin{equation}
ds^{2}=-A(r)dt^{2}+B(r)dr^{2}+C(r)d\Omega^{2}\label{eq:general-metric}
\end{equation}
With this metric, the only non-vanishing components of the tensor
$X_{\mu\nu}$ are the diagonal ones, namely, $\mu=\nu$. With respect
to the off-diagonal components, $\mu\neq\nu$, since $g_{\mu\nu}=0$,
the field equation requires
\begin{equation}
T_{\mu\nu}=0\ \ \ \text{if }\mu\neq\nu
\end{equation}
meaning the EMT has to be diagonal. The trace is merely a sum $T=T_{0}^{0}+T_{1}^{1}+T_{2}^{2}+T_{3}^{3}$.
Furthermore since
\begin{align}
g_{33} & =g_{22}\sin^{2}\theta\\
X_{33} & =X_{22}\sin^{2}\theta
\end{align}
one has
\begin{equation}
T_{33}=T_{22}\sin^{2}\theta\ \ \ \Longrightarrow\ \ \!T_{3}^{3}=T_{2}^{2}
\end{equation}
Therefore, the EMT must adopt the following form
\begin{equation}
T_{\mu}^{\nu}=\left(\begin{array}{cccc}
-\epsilon(r) & 0 & 0 & 0\\
0 & p_{\parallel}(r) & 0 & 0\\
0 & 0 & p_{\perp}(r) & 0\\
0 & 0 & 0 & p_{\perp}(r)
\end{array}\right)\label{eq:EMT}
\end{equation}
The only assumptions we made are the stationarity and spherical symmetry
for the metric components and the BD scalar field. The energy density
$\epsilon$, the radial pressure $p_{\parallel}$ and the tangential
pressure $p_{\perp}$ are functions of $r$. The trace is simplified
to
\begin{equation}
T=-\epsilon+p_{\parallel}+2p_{\perp}\label{eq:EMT-trace}
\end{equation}

We shall not impose any further constraints on the EMT, which can
be anisotropic.

\section{\label{sec:Standard-coord}Integrability of the $00-$field equation}

Let us start with the standard areal coordinate system
\begin{equation}
ds^{2}=-A(r)dt^{2}+B(r)dr^{2}+r^{2}d\Omega^{2}\label{eq:standard-metric}
\end{equation}
In this form, a star center exists at $r=0$, at which the metric
components, $A(r)$ and $B(r)$ and the BD field $\Phi(r)$ are regular
(including their first derivatives with respect to $r$). The scalar
field equation \eqref{eq:BD-eq-phi} yields
\begin{align}
\frac{d}{dr}\biggl(r^{2}\sqrt{\frac{A}{B}}\frac{d\Phi}{dr}\biggr)\sin\theta & =\frac{8\pi T}{2\omega+3}\sqrt{-g}
\end{align}
Multiply both sides with $dV:=dr\,d\theta\,d\varphi$, then integrate
\begin{align}
\int_{0}^{r}d\biggl(r^{2}\sqrt{\frac{A}{B}}\frac{d\Phi}{dr}\biggr)\int_{0}^{\pi}\sin\theta d\theta\int_{0}^{2\pi}d\varphi\ \ \ \nonumber \\
=\frac{8\pi}{2\omega+3}\int_{V}dV\sqrt{-g}\,T
\end{align}
with the integral domain $V$ being a ball of radius $r$, centered
at the origin.\vskip4pt

Since $A$, $B$, $\Phi$, $A'$, $\Phi'$ are finite at the star
center, $r=0$, (viz. regularity conditions), upon integration of
the left hand side, the above equation becomes
\begin{equation}
r^{2}\sqrt{\frac{A}{B}}\frac{d\Phi}{dr}=\frac{2}{2\omega+3}\int_{V}dV\sqrt{-g}\,T\label{eq:z-1}
\end{equation}

Next, the $00-$component of $X_{\mu\nu}$ can be written as
\begin{align}
X_{00} & =\Phi R_{00}+\Gamma_{00}^{1}\frac{d\Phi}{dr}\\
 & =\frac{\sqrt{A}}{2r^{2}\sqrt{B}}\frac{d}{dr}\biggl(\frac{r^{2}\Phi}{\sqrt{AB}}\frac{dA}{dr}\biggr)\\
 & =\frac{-\sin\theta}{2\sqrt{-g}g^{00}}\frac{d}{dr}\biggl(\frac{r^{2}\Phi}{\sqrt{AB}}\frac{dA}{dr}\biggr)\label{eq:X00}
\end{align}
\vskip10pt

\noindent It is imperative to note that \emph{the $X_{00}$ term is
expressible in a neat ``integrable'' form, Eq. \eqref{eq:X00}}. Consequently,
the $00-$component of the field equation \eqref{eq:BD-eq-R} yields
\begin{align}
\frac{d}{dr}\biggl(\frac{r^{2}\Phi}{\sqrt{AB}}\frac{dA}{dr}\biggr)\sin\theta & =-16\pi\,\Bigl(T_{0}^{0}-\frac{\omega+1}{2\omega+3}T\Bigr)\sqrt{-g}
\end{align}
Integrating from the star center (assuming regularity)
\begin{equation}
\frac{r^{2}\Phi}{\sqrt{AB}}\frac{dA}{dr}=-4\int_{V}dV\sqrt{-g}\,\Bigl(T_{0}^{0}-\frac{\omega+1}{2\omega+3}T\Bigr)\label{eq:z-2}
\end{equation}
\vskip12pt

The rest of this section deals with the terms in the right hand side
of Eqs. \eqref{eq:z-1} and \eqref{eq:z-2}. Consider a compact star
of finite radius $r_{*}$ and denote $V^{*}$ the domain of the star,
namely, $V^{*}$ being a ball centered at the origin with radius $r_{*}$.
The following integrals are defined for the domain $V^{*}$:\vskip-8pt

\begin{align}
E^{*} & :=\int_{V^{*}}dV\sqrt{-g}\,\epsilon\ \ =4\pi\int_{0}^{r_{*}}dr\,r^{2}\sqrt{AB}\,\epsilon;\nonumber \\
P_{\parallel}^{*} & :=\int_{V^{*}}dV\sqrt{-g}\,p_{\parallel}\,=4\pi\int_{0}^{r_{*}}dr\,r^{2}\sqrt{AB}\,p_{\parallel};\label{eq:integrals-def}\\
P_{\perp}^{*} & :=\int_{V^{*}}dV\sqrt{-g}\,p_{\perp}=4\pi\int_{0}^{r_{*}}dr\,r^{2}\sqrt{AB}\,p_{\perp}.\nonumber 
\end{align}
Note that outside the star, $\epsilon$, $p_{\parallel}$, and $p_{\perp}$
vanish. Therefore for a ball $V$ that is centered at the origin and
encloses $V^{*}$ (namely, its radius $r$ exceeds $r_{*}$), the
following identities hold:\vskip-8pt
\begin{align}
\int_{V}dV\sqrt{-g}\,\epsilon & =E^{*};\nonumber \\
\int_{V}dV\sqrt{-g}\,p_{\parallel} & =P_{\parallel}^{*};\label{eq:integrals-exterior}\\
\int_{V}dV\sqrt{-g}\,p_{\perp} & =P_{\perp}^{*}.\nonumber 
\end{align}
For a ball $V$ \emph{enclosing} $V^{*}$, the integrals in the right
hand side of Eqs. \eqref{eq:z-1} and \eqref{eq:z-2} are
\begin{align}
\int_{V}dV\sqrt{-g}\,T & =\int_{V}dV\sqrt{-g}\left[-\epsilon+p_{\parallel}+2p_{\perp}\right]\\
 & =-E^{*}+P_{\parallel}^{*}+2P_{\perp}^{*}\label{eq:z-3}
\end{align}
and
\begin{align}
 & \int_{V}dV\sqrt{-g}\Bigl(T_{0}^{0}-\frac{\omega+1}{2\omega+3}T\Bigr)\nonumber \\
 & =\int_{V}dV\sqrt{-g}\Bigl[-\frac{\omega+2}{2\omega+3}\epsilon-\frac{\omega+1}{2\omega+3}\bigl(p_{\parallel}+2p_{\perp}\bigr)\Bigr]\\
 & =-\frac{\omega+2}{2\omega+3}E^{*}-\frac{\omega+1}{2\omega+3}\bigl(P_{\parallel}^{*}+2P_{\perp}^{*}\bigr)\label{eq:z-4}
\end{align}
\vskip8pt
\begin{rem}
The integrability property that we exploited in deriving Eqs. \eqref{eq:z-1}
and \eqref{eq:z-2} is not limited to the standard coordinates. If
we used the metric in Eq. \eqref{eq:general-metric}, the left hand
side of Eqs. \eqref{eq:z-1} and \eqref{eq:z-2} would read $C\sqrt{\frac{A}{B}}\frac{d\Phi}{dr}$
and $\frac{C\Phi}{\sqrt{AB}}\frac{dA}{ar}$ respectively, and the
$\sqrt{-g}$ term in the right hand side of these equation would be
calculated using the latter metric. Nevertheless, the standard-coordinate
metric in \eqref{eq:standard-metric} explicitly reveals the presence
of the point $r=0$ at which the surface area of a 2-sphere vanishes.
This point naturally corresponds to the star's center and serves as
the lower bound in the integrals defined in \eqref{eq:integrals-def}.
\end{rem}
\vskip6pt
\begin{rem}
In the integrals defined in Eq. \eqref{eq:integrals-def}, the element
$dV\sqrt{-g}$ is equal to $\sqrt{A}\left(r^{2}\sqrt{B}dr\,\sin\theta d\theta\,d\varphi\right)$.
The term $\sqrt{A}$, equal to $\sqrt{-g_{00}}$, is a ``redshift
factor''. The combination $r^{2}\sqrt{B}dr\,\sin\theta d\theta\,d\varphi$,
equal to $r^{2}\sqrt{g_{11}}dr\,\sin\theta d\theta\,d\varphi$, is
the spatial volume element in the spatial part of the metric given
in Eq. \eqref{eq:standard-metric}. Note that in the region occupied
with matter, e.g. where $\epsilon(r)\neq0$, the space in general
is not Euclidean; consequently, for the standard-coordinate metric,
$B(r)$ deviates from $1$.\vskip6pt

It should also be noted that the combination $r^{2}\sqrt{g_{11}}dr\,\sin\theta d\theta\,d\varphi$
is equal to $\sqrt{g^{(3)}}dV$, where $g^{(3)}$ is the determinant
of the spatial section of the metric in Eq. \eqref{eq:standard-metric}.
The combination is thus \emph{invariant} with respect to a twice-differentiable
transformation of the radial coordinate. The quantities $E^{*}$,
$P_{\parallel}^{*}$, and $P_{\perp}^{*}$ can be interpreted as the
total energy, radial pressure, and tangential pressure contained within
the compact star.
\end{rem}

\section{\textcompwordmark\label{sec:Isotropic-coord}Transforming to isotropic
coordinates}

With the total energy and pressures defined via Eq. \eqref{eq:integrals-exterior},
the set of equations \eqref{eq:z-1}, \eqref{eq:z-2}, \eqref{eq:z-3},
and \eqref{eq:z-4} essentially provide the ``conservation'' rules
for the metric components and the BD scalar in the \emph{exterior}
vacuum, i.e., for $r>r_{*}$, per
\begin{equation}
r^{2}\sqrt{\frac{A}{B}}\frac{d\Phi}{dr}=\frac{2}{2\omega+3}\left[-E^{*}+P_{\parallel}^{*}+2P_{\perp}^{*}\right]\label{eq:a-1}
\end{equation}
and
\begin{equation}
\frac{r^{2}\Phi}{\sqrt{AB}}\frac{dA}{dr}=\frac{4}{2\omega+3}\left[(\omega+2)E^{*}+(\omega+1)\bigl(P_{\parallel}^{*}+2P_{\perp}^{*}\bigr)\right]\label{eq:a-2}
\end{equation}
Note that the right hand sides of Eqs. \eqref{eq:a-1} and \eqref{eq:a-2}
are integration \emph{constants}, induced by the matter distribution
in the interior of the compact mass source.\vskip4pt

Our next step is to relate the \emph{parameters} of the exterior vacuum
to these constants. The vacuum solution for BD gravity is best known
in the \emph{isotropic} coordinate, in the form of the Brans Class
I solution. (The issue with the generality and uniqueness of the Brans
Class I solution shall be addressed in Section \ref{sec:Unique}.)
It is thus necessary to bring the left hand sides of Eqs. \eqref{eq:a-1}
and \eqref{eq:a-2} to the isotropic coordinate $\rho$, namely, using
the following metric

\begin{align}
ds^{2} & =-F(\rho)dt^{2}+G(\rho)\left(d\rho^{2}+\rho^{2}d\Omega^{2}\right)\label{eq:isotropic-metric}
\end{align}
and the BD scalar $\phi(\rho)$. Transforming \eqref{eq:standard-metric}
into \eqref{eq:isotropic-metric} requires the following identifications
\begin{equation}
F(\rho)=A(r);\ \ G(\rho)\left(\frac{d\rho}{dr}\right)^{2}=B(r);\ \ G(\rho)\,\rho^{2}=r^{2}\label{eq:ident}
\end{equation}
in addition to a mapping for the BD scalar
\begin{equation}
\phi(\rho)=\Phi(r)
\end{equation}
In the far-field region, it is expected that the relation between
$r$ and $\rho$ is monotonic, viz. $\frac{d\rho}{dr}>0$. The quantities
of interest thence become
\begin{align}
r^{2}\sqrt{\frac{A}{B}}\frac{d\Phi}{dr} & =G\rho^{2}\sqrt{\frac{F}{G\left(\frac{d\rho}{dr}\right)^{2}}}\frac{d\phi}{d\rho}\frac{d\rho}{dr}=\rho^{2}\sqrt{FG}\frac{d\phi}{d\rho}
\end{align}
and
\begin{align}
\frac{r^{2}\Phi}{\sqrt{AB}}\frac{dA}{dr} & =\frac{G\rho^{2}\phi}{\sqrt{FG\left(\frac{d\rho}{dr}\right)^{2}}}\frac{dF}{d\rho}\frac{d\rho}{dr}=\rho^{2}\phi\sqrt{\frac{G}{F}}\frac{dF}{d\rho}
\end{align}
We thus arrive at
\begin{equation}
\rho^{2}\sqrt{FG}\frac{d\phi}{d\rho}=\frac{2}{2\omega+3}\left[-E^{*}+P_{\parallel}^{*}+2P_{\perp}^{*}\right]\label{eq:a-3}
\end{equation}
and
\begin{equation}
\rho^{2}\phi\sqrt{\frac{G}{F}}\frac{dF}{d\rho}=\frac{4}{2\omega+3}\left[(\omega+2)E^{*}+(\omega+1)\bigl(P_{\parallel}^{*}+2P_{\perp}^{*}\bigr)\right]\label{eq:a-4}
\end{equation}
\vskip8pt
\begin{rem}
As we stated at the end of the preceding section, the energy density
and pressure quantities are invariant upon a coordinate transformation
in the radial coordinate. This can be verified for the case at hand.
For example, the total energy in the standard coordinates is
\begin{align}
E_{\text{std}}^{*} & =\int_{V}dr\,d\theta\,d\varphi\sqrt{A(r)B(r)}\,r^{2}\sin\theta\,\epsilon(r)
\end{align}
whereas in the isotropic coordinates
\begin{align}
E_{\text{iso}}^{*} & =\int_{V}d\rho\,d\theta\,d\varphi\sqrt{F(\rho)G^{3}(\rho)}\,\rho^{2}\sin\theta\,\epsilon'(\rho)
\end{align}
with the identification $\epsilon'(\rho)=\epsilon\left(r(\rho)\right)$.
It is straightforward to see that 
\begin{align}
E_{\text{std}}^{*} & =\int_{V}\left(d\rho\frac{dr}{d\rho}\right)d\theta d\varphi\sqrt{FG\left(\frac{d\rho}{dr}\right)^{2}}G\rho^{2}\sin\theta\,\epsilon\left(r(\rho)\right)\\
 & =\int_{V}d\rho\,d\theta\,d\varphi\sqrt{FG^{3}}\,\rho^{2}\sin\theta\,\epsilon'(\rho)\\
 & =E_{\text{iso}}^{*}
\end{align}
Hence, the energy integral $E^{*}$ is the same for both systems---the
standard and the isotropic coordinates. Likewise, the same conclusion
applies for the pressure integrals $P_{\parallel}^{*}$ and $P_{\perp}^{*}$.
\end{rem}

\section{\label{sec:Unique}Brans Class I as the unique vacuum solution for
$\omega>-3/2$}

Equations \eqref{eq:a-3} and \eqref{eq:a-4} stand handy for us to
relate the parameters of the exterior vacuum solution to the energy-pressure
integrals. It is well documented that the Brans Class I solution is
a vacuum solution in BD gravity. In this section, we shall \emph{further}
demonstrate that, for the case of non-phantom kinetic energy for the
BD field, viz. $\omega>-3/2$, the Brans Class I solution is the most
general \emph{and} unique vacuum solution. That is to say, the Brans
Class I solution is \emph{the} vacuum solution in BD gravity, when
$\omega>-3/2$.\vskip4pt

Historically, Brans discovered the solutions during his PhD thesis
and reported them in \citep{Brans-1962} without offering a derivation,
although the solutions can be verified via direct inspection. The
earliest public account for an analytical derivation of these solutions
can be traced back to Ref. \citep{Bronnikov-1973} in which Bronnikov
discovered the more general Brans-Dicke-Maxwell electrovacuum solution.\vskip4pt

The Brans solutions are comprised of 4 different classes (or types).
In the exposition of Bronnikov \citep{Bronnikov-1973}, the Brans-Dicke
theory is first mapped from the Jordan frame to the Einstein frame
via a Weyl mapping. The transformed BD scalar field becomes uncoupled
(thus free) scalar field which sources the Einstein-frame spacetime
metric. This transformed theory is known to admit the Fisher--Janis--Newman--Winicour
(FJNW) solution. (Note: The solution has been re-discovered several
times, and is also known as the Fisher--Bergmann--Leipnik--\linebreak
Janis--Newman--Winicour--Buchdahl--Wyman (FBLJNWBW) solution \citep{Faraoni-2018,Wyman-1981,Bergmann-1957,Buchdahl-BD-1972,Fisher-1948,Janis-1968}.)
Via the Einstein-frame representation, it has been established by
now that the classes of Brans solutions are the \emph{most general}
solutions that are static and spherisymmetric in Brans-Dicke gravity
\citep{Bhadra-2005,Faraoni-2018}.\vskip4pt

However, the existence of multiple Brans classes has sidetracked a
recognition of their ``uniqueness''. There is a degree of redundancy
in these classes, however. In \citep{Bhadra-2005} Bhadra and Sarkar
pointed out that Class III and Class IV are equivalent via a coordinate
transformation, $\rho\leftrightarrow1/\rho$, reducing the count of
classes to 3. These authors also uncovered a ``symmetry'' in terms
of \emph{parameters} of Brans Class I and Brans Class II, upon making
certain replacement in the parameters (and a coordinate transform)
\footnote{The replacement of parameters to map Class II into Class I can also
be viewed as a ``Wick rotation'' \citep{Nandi-2010}.}. Consequently, it was deemed that the two classes, I and II, were
equivalent, leaving only Class I and Class III to be truly independent.
Nomenclature aside, it should be noted that diffeomorphism alone cannot
transform Class I into Class II and vice versa. In \citep{Faraoni-2018}
Faraoni and colleagues revisited this issue and correctly branded
the ``symmetry'' alluded above a ``duality'' rather than an ``equivalence''.
As we shall show momentarily, the 3 Brans classes (I, II, and IV)
remain \emph{separate} solutions (while fully covering) in the parameter
space. Yet, the 3 classes can be ``unified'' upon an appropriate parametrization.
That is to say, all 3 Brans classes of solution can be brought into
a single form, as we shall do below.\vskip8pt

Consider the metric and scalar field
\begin{equation}
\left\{ \begin{array}{rl}
ds^{2} & ={\displaystyle -F(\rho)dt^{2}+G(\rho)\left(d\rho^{2}+\rho^{2}d\Omega^{2}\right)}\\
\\
F(\rho) & {\displaystyle =\left(\frac{\rho-\frac{1}{2}M_{1}\sqrt{\kappa}}{\rho+\frac{1}{2}M_{1}\sqrt{\kappa}}\right)^{\frac{2}{\sqrt{\kappa}}}}\\
G(\rho) & ={\displaystyle \left(1-\frac{M_{1}^{2}\kappa}{4\rho^{2}}\right)^{2}\left(\frac{\rho-\frac{1}{2}M_{1}\sqrt{\kappa}}{\rho+\frac{1}{2}M_{1}\sqrt{\kappa}}\right)^{-\frac{2(1+\Lambda)}{\sqrt{\kappa}}}}\\
\phi(\rho) & ={\displaystyle \left(\frac{\rho-\frac{1}{2}M_{1}\sqrt{\kappa}}{\rho+\frac{1}{2}M_{1}\sqrt{\kappa}}\right)^{\frac{\Lambda}{\sqrt{\kappa}}}}
\end{array}\right.\label{eq:unify}
\end{equation}
with $M_{1}\in\mathbb{R}^{+}$ (a parameter of length dimension) and
dimensionless parameters $\Lambda\in\mathbb{R}$ and $\kappa\in\mathbb{R}$.
It is straightforward to verify by direct inspection that the metric
and scalar field satisfy the BD field equations, with the parameter
$\kappa\in\mathbb{R}$ obeying ($\omega\in\mathbb{R}$ and $\Lambda\in\mathbb{R}$)
\begin{align}
\kappa & :=(1+\Lambda)^{2}-\Lambda\bigl(1-\frac{\omega}{2}\Lambda\bigr)\label{eq:kappa}
\end{align}
Three cases to consider, depending on the signum of $\kappa$:\vskip12pt
\begin{figure}[!t]
\begin{centering}
\includegraphics[scale=0.75]{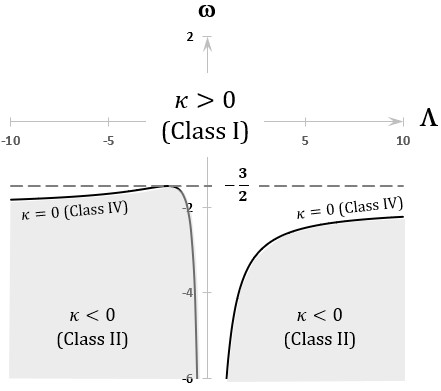}\vspace{.25cm}
\par\end{centering}
\caption{\label{fig:Parameter-space-Brans}Parameter space for the ``unified''
Brans solution, Eq. \eqref{eq:unify}. The solid line corresponds
to $(1+\Lambda)^{2}-\Lambda\bigl(1-\omega\Lambda/2\bigr)=0$. The
two branches of $\kappa=0$ asymptote $\omega\rightarrow-2$ at large
values of $\Lambda$. The local peak in the left branch occurs at
\{$\omega=-3/2,\,\Lambda=-2\}$. The region with $\omega>-3/2$ is
Brans Class I, exclusively.}
\end{figure}

Case 1:$\ $ For $\kappa>0$, it is evidently Brans Class I solution
as originally reported in \citep{Brans-1962}.\vskip12pt

Case 2:$\ $ For $\kappa<0$, utilizing the identity $\tan^{-1}x=\frac{i}{2}\ln\frac{1-i\,x}{1+i\,x}$,
one has
\begin{equation}
\frac{1}{\sqrt{\kappa}}\ln\frac{\rho-\frac{1}{2}M_{1}\sqrt{\kappa}}{\rho+\frac{1}{2}M_{1}\sqrt{\kappa}}=\frac{-2}{\sqrt{-\kappa}}\tan^{-1}\frac{M_{1}\sqrt{-\kappa}}{2\rho}
\end{equation}
The change of variable $\rho=\frac{1}{4}M_{1}^{2}\kappa/\rho'$, hence
$d\rho=-\frac{1}{4}M_{1}^{2}\kappa/\rho'^{2}d\rho'$, renders
\begin{align}
\left(\frac{\rho-\frac{1}{2}M_{1}\sqrt{\kappa}}{\rho+\frac{1}{2}M_{1}\sqrt{\kappa}}\right)^{\frac{1}{\sqrt{\kappa}}} & =e^{\frac{2}{\sqrt{-\kappa}}\tan^{-1}\frac{2\rho'}{M_{1}\sqrt{-\kappa}}}\\
1-\frac{M_{1}^{2}\kappa}{4\rho^{2}} & =1+\frac{4\rho'^{2}}{M_{1}^{2}(-\kappa)}
\end{align}
resulting in Brans Class II solution in the new radial coordinate
$\rho'$ \citep{Brans-1962}:
\begin{align}
F(\rho') & =e^{\frac{4}{\sqrt{-\kappa}}\tan^{-1}\frac{2\rho'}{M_{1}\sqrt{-\kappa}}}\\
G(\rho') & =\left(1+\frac{M_{1}^{2}(-\kappa)}{4\rho'^{2}}\right)^{2}e^{-\frac{2(1+\Lambda)}{\sqrt{-\kappa}}\tan^{-1}\frac{2\rho'}{M_{1}\sqrt{-\kappa}}}\\
\phi(\rho') & =e^{\frac{2\Lambda}{\sqrt{-\kappa}}\tan^{-1}\frac{2\rho'}{M_{1}\sqrt{-\kappa}}}.
\end{align}
\vskip8pt

Case 3:$\ $ For $\kappa=0$, the limit
\begin{equation}
\left(\frac{\rho-\frac{1}{2}M_{1}\sqrt{\kappa}}{\rho+\frac{1}{2}M_{1}\sqrt{\kappa}}\right)^{\frac{1}{\sqrt{\kappa}}}\underset{\kappa\rightarrow0}{\simeq}e^{-\frac{M_{1}}{\rho}}
\end{equation}
produces Brans Class IV solution \citep{Brans-1962}: \footnote{Brans Class III is obtainable from Brans Class IV via a coordinate
transform, $\rho=M_{1}^{2}/(4\rho')$.}
\begin{align}
F(\rho) & =e^{-\frac{M_{1}}{\rho}}\\
G(\rho) & =e^{\frac{(1+\Lambda)M_{1}}{\rho}}\\
\phi(\rho) & =e^{-\frac{\Lambda M_{1}}{\rho}}.
\end{align}
\vskip8pt

Let us emphasize that the recasting exercise thus far represents \emph{nothing
essentially new}. The important point which our recasting offers is
that the signum of $\kappa$ elects the ``class'' that the ``unified''
Brans solution, Eq. \eqref{eq:unify}, belongs. Moreover, a pair of
$\{\omega,\,\Lambda\}$ \emph{uniquely determines the signum for $\kappa$},
and hence the form of the solution. Figure \ref{fig:Parameter-space-Brans}
shows the categorization. In the $\{\omega,\,\Lambda\}$ plane, the
two branches of the solid line correspond to the loci of $\kappa=0$,
viz. $(1+\Lambda)^{2}-\Lambda\bigl(1-\omega\Lambda/2\bigr)=0$. Above
and below the solid line are the domains for Brans Class I and Brans
Class II, respectively. The 3 classes do not overlap while fully covering
the $\{\omega,\,\Lambda\}$ plane. The ambiguity in selecting the
solution is removed. The \emph{uniqueness} of the ``unified'' Brans
solution, Eq. \eqref{eq:unify}, is thereby established.\vskip12pt
\begin{rem}
For $\omega>-3/2$, the function $\kappa$ is positive-definite for
all values of $\Lambda$ in $\mathbb{R}$, as can be seen by rewriting
$\kappa$ as
\begin{align}
\kappa & =\frac{\left[(\omega+2)\Lambda+1\right]^{2}+2\omega+3}{2(\omega+2)}>0\ \ \ \text{for }\omega>-3/2
\end{align}
Consequently, only Brans Class I is admissible for the non-phantom
action, $\omega>-3/2$, a fact evident in Fig. \ref{fig:Parameter-space-Brans}.
\end{rem}
\vskip6pt
\begin{rem}
It is worth noting that Brans Class I recovers the Schwarzschild metric
when $\Lambda=0$ forcing $\kappa=1$. On the other hand, Brans Class
II and Class III do not have this property and are often relegated
as ``pathological'' solutions. Conveniently, they only exist for the
phantom action, i.e., $\omega\leqslant-3/2$.
\end{rem}
\vskip6pt

Another way to recognize the unified nature of the 3 Brans solutions
is via the harmonic coordinate instead of the isotropic coordinate
\citep{Bronnikov-1973}. For completeness, we shall revisit this representation
in Appendix \ref{sec:Mapping-Brans-solutions}.

\section{\label{sec:Matching}Matching the exterior solution with energy-pressure
integrals}

As established in the preceding section, for the non-phantom kinetic
action (i.e., $\omega>-3/2$), the exterior vacuum solution is \emph{exclusively}
Brans Class I. We are now equipped to perform the matching of its
parameters.\vskip4pt

With the metric given in Eq. \eqref{eq:unify}, the left hand sides
of Eqs. \eqref{eq:a-3} and \eqref{eq:a-4} readily yield
\begin{align}
\rho^{2}\sqrt{FG}\,\frac{d\phi}{d\rho} & =M_{1}\Lambda\\
\rho^{2}\phi\sqrt{\frac{G}{F}}\,\frac{dF}{d\rho} & =2M_{1}
\end{align}
With the aid of Eqs. \eqref{eq:a-3} and \eqref{eq:a-4}, we deduce
that
\begin{align}
M_{1} & =2\,\frac{(\omega+2)E^{*}+(\omega+1)\bigl(P_{\parallel}^{*}+2P_{\perp}^{*}\bigr)}{2\omega+3}\\
\Lambda & =\frac{-E^{*}+P_{\parallel}^{*}+2P_{\perp}^{*}}{(\omega+2)E^{*}+(\omega+1)\bigl(P_{\parallel}^{*}+2P_{\perp}^{*}\bigr)}
\end{align}
Denote the dimensionless ratio
\begin{equation}
\Theta:=\frac{P_{\parallel}^{*}+2P_{\perp}^{*}}{E^{*}}\label{eq:varTheta-def}
\end{equation}
For an isotropic EMT, $P_{\parallel}^{*}=P_{\perp}^{*}\equiv P^{*}$,
the quantity $\Theta$ is $3P^{*}/E^{*}$. In general, thence
\begin{align}
M_{1} & =2E^{*}\,\frac{(\omega+2)+(\omega+1)\,\Theta}{2\omega+3}\label{eq:M1-formula}\\
\Lambda & =\frac{\Theta-1}{(\omega+2)+(\omega+1)\,\Theta}\label{eq:Lambda-formula}
\end{align}
Note that for $\omega>-3/2$, $E>0$, and $0\leqslant\Theta<1$, the
positive-definiteness of $M_{1}$ is ensured
\begin{equation}
M_{1}=E^{*}\left[1+\Theta+\frac{1-\Theta}{2\omega+3}\right]\geqslant E^{*}.
\end{equation}
Furthermore, from Eq. \eqref{eq:kappa}, the parameter $\kappa$
\begin{align}
\kappa & =\frac{2\omega+3}{4}\,\frac{(2\omega+3)\left(1+\Theta\right)^{2}+\left(1-\Theta\right)^{2}}{\left((\omega+2)+(\omega+1)\,\Theta\right)^{2}}
\end{align}
giving $\kappa>0$ for $\omega>-3/2$, confirming the validity of
the Brans Class I solution under consideration.

\section{\label{sec:Robertson-params}The Robertson parameters}

The Robertson expansion in isotropic coordinates is \citep{Weinberg}:
\begin{align}
ds^{2} & =-\left(1-2\,\frac{M_{1}}{\rho}+2\beta\,\frac{M_{1}^{2}}{\rho^{2}}+\dots\right)dt^{2}\nonumber \\
 & \ \ \ \,+\left(1+2\gamma\,\frac{M_{1}}{\rho}+\dots\right)\left(d\rho^{2}+\rho^{2}d\Omega^{2}\right)\label{eq:Weinberg}
\end{align}
in which $\beta$ and $\gamma$ are the Robertson (or Eddington-Robertson-Schiff)
parameters. The metric in Eq. \eqref{eq:unify} can be re-expressed
in the expansion form
\begin{align}
ds^{2} & =-\left(1-2\,\frac{M_{1}}{\rho}+2\,\frac{M_{1}^{2}}{\rho^{2}}+\dots\right)dt^{2}\nonumber \\
 & +\left(1+2\left(1+\Lambda\right)\frac{M_{1}}{\rho}+\dots\right)\left(d\rho^{2}+\rho^{2}d\Omega^{2}\right)\label{eq:Robertson}
\end{align}
Comparing Eqs. \eqref{eq:Weinberg} against Eq. \eqref{eq:Robertson},
we obtain
\begin{align}
\beta_{\,\text{exact}} & \ =\ 1\label{eq:beta-exact-0}\\
\gamma_{\,\text{exact}} & \ =\ 1+\Lambda\label{eq:gamma-exact-0}
\end{align}
where we have added in the subscript ``exact'' as emphasis. Note that
$\Lambda$ measures the deviation of the $\gamma$ parameters from
GR ($\gamma_{\text{GR}}=1$). From Eq. \eqref{eq:Lambda-formula},
\emph{$\Lambda$ depends on both $\omega$ and $\Theta$}. Finally,
we arrive at
\begin{align}
\gamma_{\,\text{exact}}\  & =\ \frac{\omega+1+(\omega+2)\,\Theta}{\omega+2+(\omega+1)\,\Theta}\label{eq:gamma-exact-1}
\end{align}
which can also be conveniently recast as
\begin{align}
\gamma_{\,\text{exact}}\  & =\ \frac{\gamma_{\,\text{PPN}}+\Theta}{1+\gamma_{\,\text{PPN}}\,\Theta}\label{eq:gamma-exact-2}
\end{align}
by recalling that $\gamma_{\,\text{PPN}}=\frac{\omega+1}{\omega+2}$.
With the aid of Eqs. \eqref{eq:varTheta-def} and \eqref{eq:M1-formula},
the active gravitational mass is
\begin{align}
M_{\text{grav}} & :=M_{1}\\
 & =E^{*}+P_{\parallel}^{*}+2P_{\perp}^{*}+\frac{E^{*}-\bigl(P_{\parallel}^{*}+2P_{\perp}^{*}\bigr)}{2\omega+3}\label{eq:active-mass}
\end{align}
where the contribution of pressure to the active gravitational mass
is evident \citep{Baez-2005,Ehlers-2005}.

\section{\label{sec:Tolman-mass}Mass relations in Brans-Dicke gravity}

In GR, the Tolman mass was defined as \citep{Tolman-1930,Florides-2009,Vollick-2021}
\begin{align}
m_{\text{T}} & :=\int_{V}dV\sqrt{-g}\left(-T_{0}^{0}+T_{1}^{1}+T_{2}^{2}+T_{3}^{3}\right)
\end{align}
In our EMT form \eqref{eq:EMT}, this renders
\begin{align}
m_{\text{T}} & =\int_{V}dV\sqrt{-g}\left(\,\epsilon+p_{\parallel}+2p_{\perp}\right)\\
 & =E^{*}+P_{\parallel}^{*}+2P_{\perp}^{*}
\end{align}
For a metric that has the following asymptotic form (i.e., as $\rho\rightarrow\infty$)
\citep{Vollick-2021}:
\begin{align}
ds^{2} & =-\left(1-\frac{2M_{\text{grav}}}{\rho}\right)dt^{2}\nonumber \\
 & +\left(1+\frac{2M_{\text{ADM}}}{\rho}\right)\left(d\rho^{2}+\rho^{2}d\Omega^{2}\right)
\end{align}
the quantity $M_{\text{grav}}$ is the active gravitation mass of
the source, whereas $M_{\text{ADM}}$ is the ADM mass. In GR, it is
known that
\begin{equation}
M_{\text{grav}}=M_{\text{ADM}}=M_{\text{T}}
\end{equation}

For BD gravity, besides Expression \eqref{eq:active-mass} for $M_{\text{grav}}$,
viz.
\begin{equation}
M_{\text{grav}}=M_{\text{T}}+\frac{E^{*}-\bigl(P_{\parallel}^{*}+2P_{\perp}^{*}\bigr)}{2\omega+3}
\end{equation}
we can also calculate the ADM mass, with the aid of Eqs. \eqref{eq:Weinberg}
and \eqref{eq:gamma-exact-1}
\begin{align}
M_{\text{ADM}} & =\gamma\,M_{\text{grav}}\\
 & =E^{*}+P_{\parallel}^{*}+2P_{\perp}^{*}-\frac{E^{*}-\bigl(P_{\parallel}^{*}+2P_{\perp}^{*}\bigr)}{2\omega+3}\\
 & =M_{\text{T}}-\frac{E^{*}-\bigl(P_{\parallel}^{*}+2P_{\perp}^{*}\bigr)}{2\omega+3}
\end{align}
The difference
\begin{equation}
M_{\text{grav}}-M_{\text{ADM}}=\frac{E^{*}-\bigl(P_{\parallel}^{*}+2P_{\perp}^{*}\bigr)}{\omega+3/2}\geqslant0
\end{equation}
for $\omega>-3/2$ and normal matter, viz. $p_{\parallel}\leqslant\frac{1}{3}\epsilon$
and $p_{\perp}^{*}\leqslant\frac{1}{3}\epsilon$. It is interesting
to note that although $M_{\text{grav}}$, $M_{\text{ADM}}$, and $M_{\text{T}}$
have different values in BD gravity, the mean value of $M_{\text{grav}}$
and $M_{\text{ADM}}$ precisely equals $M_{\text{T}}$.\vskip12pt
\begin{rem}
In the limit of infinite $\omega$, we recover the usual relation
in GR
\begin{equation}
m_{\text{grav}}=m_{\text{ADM}}=m_{\text{T}}=E^{*}+P_{\parallel}^{*}+2P_{\perp}^{*}\label{eq:mass-in-GR}
\end{equation}
Furthermore, in GR, the ADM mass has been established within the Tolman-Oppenheimer-Volkoff
framework (using the standard coordinates) to be
\begin{equation}
m_{\text{ADM}}=4\pi\int_{0}^{r_{*}}dr\,r^{2}\,\epsilon(r)
\end{equation}
Using Eq. \eqref{eq:integrals-def} in the standard coordinates \eqref{eq:standard-metric},
we rewrite the right hand side of Eq. \eqref{eq:mass-in-GR} as
\begin{equation}
E^{*}+P_{\parallel}^{*}+2P_{\perp}^{*}=4\pi\int_{0}^{r_{*}}dr\,r^{2}\sqrt{-g_{00}\,g_{11}}\left(\,\epsilon+p_{\parallel}+2p_{\perp}\right)
\end{equation}
Combining the last 3 equations yields
\begin{equation}
\int_{0}^{r_{*}}dr\,r^{2}\,\epsilon\ =\int_{0}^{r_{*}}dr\,r^{2}\sqrt{-g_{00}\,g_{11}}\left(\,\epsilon+p_{\parallel}+2p_{\perp}\right)
\end{equation}
We thus have re-derived the Tolman relation in GR, as a by-product
of our study \citep{Tolman-1930,Vollick-2021}.
\end{rem}

\section{\label{sec:Higher-PPN}Extension to other PPN levels}

In principle, once the Brans Class I solution for
the exterior vacuum is \emph{fully} determined by the energy and pressure integrals, \emph{all} PN parameters (of any PPN
level) can be readily calculated. These PN parameters would be useful
for estimating the values of $\omega$ and $\Theta$, since a sole
measurement of $\gamma$ is insufficient to fix \emph{two} unknowns. In this section, we illustrate the utility of the complete
Brans Class I for a second-order PN parameter $\delta$. It is worth
noting that ongoing and future space missions, such as the Gaia Mission,
require the second-order for light propagation in some specific situations
\citep{2PN}, thereby underscoring the importance of deriving $\delta$
in this context.\vskip4pt

The 2PN extension in the isotropic coordinate system is
\begin{equation}
g_{\rho\rho}=1+2\gamma\,\frac{M_{1}}{\rho}+\frac{3}{2}\delta\,\frac{M_{1}^{2}}{\rho^{2}}+\dots
\end{equation}
Note that for Schwarzschild metric, the expansion above yields $\gamma=\delta=1$.
The expansion of the Brans Class I gives
\begin{equation}
g_{\rho\rho}=1+2\left(1+\Lambda\right)\frac{M_{1}}{\rho}+\left[2\left(1+\Lambda\right)^{2}-\frac{1}{2}\kappa\right]\frac{M_{1}^{2}}{\rho^{2}}+\dots
\end{equation}
with $\kappa$ given in Eq. \eqref{eq:kappa}. Taken together and
in conjunction with Eq. \eqref{eq:Lambda-formula}, these expansions
render
\begin{align}
\delta_{\,\tiny\text{exact}} & =\ 1+\frac{7}{3}\Lambda+\left(1-\frac{\omega}{6}\right)\Lambda^{2}\\
 & =\frac{1}{\left[\omega+2+(\omega+1)\,\Theta\right]^{2}}\Bigl[\Bigl(\omega^{2}+\frac{3}{2}\omega+\frac{1}{3}\Bigr)\nonumber \\
 & +\Bigl(2\omega^{2}+\frac{19}{3}\omega+\frac{13}{3}\Bigr)\,\Theta+\Bigl(\omega^{2}+\frac{25}{6}\omega+\frac{13}{3}\Bigr)\,\Theta^{2}\Bigr]\label{eq:delta-exact-1}
\end{align}
When $\omega\rightarrow\infty$, this formula recovers the GR value
\begin{equation}
\delta_{\,\tiny\text{exact}}\rightarrow1\,.
\end{equation}
It should also be noted that, for ultra-relativistic matter, viz.
$\Theta\rightarrow1^{-}$, as $\Lambda\rightarrow0$, the formula
gives
\begin{equation}
\delta_{\,\tiny\text{exact}}\rightarrow1\,.
\end{equation}

\section{\label{sec:Discussions}Discussions}

Formula \eqref{eq:gamma-exact-1} is one of the most important outcomes
of our paper. This section aims to clarify a number of logical and
technical steps taken in the derivation, and discusses the implications
of Formula \eqref{eq:gamma-exact-1}. \vskip8pt

Our derivation proceeded in the following steps:
\begin{enumerate}
\item Assuming the metric and the mass source to be stationary and spherically
symmetric, we deduced from the BD field equation that the most general
EMT of the source can be put in the form $T_{\mu}^{\nu}=\text{diag}\left(-\epsilon(r),\,p_{\parallel}(r),\,p_{\perp}(r),\,p_{\perp}(r)\right)$.
Whereas the EMT can be anisotropic, we imposed no further conditions
on the EMT, such as being a perfect fluid. See Section \ref{sec:EMT}.
\item The standard coordinate system, $ds^{2}=-A(r)dt^{2}+B(r)dr^{2}+r^{2}d\Omega^{2}$,
is suitable for describing stars. Using the scalar field equation
for $\Phi$ and the $00-$component of the field equation, and imposing
regularity conditions at the star's center, we derived two ODE's---Eqs.
\eqref{eq:z-1} and \eqref{eq:z-2}---which express $A(r)$, $B(r)$
and $\Phi(r)$ in the \emph{exterior vacuum} in terms of the total
energy $E^{*}$ and total pressure $P_{\parallel}^{*}+2P_{\perp}^{*}$
\emph{contained within the star}. The standard coordinate system allows
for the existence of the point $r=0$ at which the surface area of
the 2-sphere vanishes, which naturally serves as the star's center.
This point acts as the lower bound for the domain of integration for
$E^{*}$ and $P_{\parallel}^{*}+2P_{\perp}^{*}$. See Section \ref{sec:Standard-coord}.
\item We next transformed the two aforementioned ODE's into the isotropic
coordinate system, $ds^{2}=-F(\rho)dt^{2}+G(\rho)\left(d\rho^{2}+\rho^{2}d\Omega^{2}\right)$.
The resulting ODE's in terms of $F(\rho)$, $G(\rho)$ and $\phi(\rho)$
are Eqs. \eqref{eq:a-3} and \eqref{eq:a-4}. The advantage of the
isotropic coordinates is that the exterior solution is best known
in this system (i.e., the Brans solutions). Furthermore, the energy
and pressure integrals, $E^{*}$ and $P_{\parallel}^{*}+2P_{\perp}^{*}$,
are unchanged when moving to this system. See Section \ref{sec:Isotropic-coord}.
\item We next presented the ``unified'' Brans solution, Eq. \eqref{eq:unify},
which cover all Brans Classes I, II and IV (with Class III being equivalent
to Class IV). For a non-phantom action, i.e. $\omega>-3/2$, Brans
Class I is the most general and \emph{unique} static spherisymmetric
vacuum solution. For a phantom action, i.e. $\omega<-3/2$, all 3
Brans classes are admissible; however, only one single class is selected
for a given set of parameters of the ``unified'' Brans solution. The
uniqueness of the ``unified'' Brans solution is thus established.
See Section \ref{sec:Unique}.
\item We employed \emph{the} Brans Class I solution to perform the matching
of the functions $F(\rho)$, $G(\rho)$ and $\phi(\rho)$ with the
energy and pressure integrals, $E^{*}$ and $P_{\parallel}^{*}+2P_{\perp}^{*}$.
See Section \ref{sec:Matching}.
\item Expressing the Brans Class I metric in the Robertson expansion, we
obtain the $\gamma$ and $\delta$ PN parameters in terms of $\omega$
and a (dimensionless) $\Theta$, defined as the ratio of $P_{\parallel}^{*}+2P_{\perp}^{*}$
and $E^{*}$. See Sections \ref{sec:Robertson-params} and \ref{sec:Higher-PPN}.
\item As a by-product, we obtained two mathematical relations linking the
active gravitational mass and the ADM mass with the energy and pressure
integrals for BD gravity. In addition, we (re)-derived the Tolman
relation in GR. See Section \ref{sec:Tolman-mass}.
\end{enumerate}
\vskip4pt

\emph{Generality of our derivation}---(i)\emph{$\ $Non-perturbative
approach}: Our derivation is non-perturbative in nature. It makes
use of the \emph{integrability} of the $00-$component of the BD field
equation, along with the scalar field equation involving $\square\,\Phi$.
(ii)\emph{$\ $Minimal assumptions}: The physical assumptions employed
are the regularity at the star's center and the existence of the star's
surface separating the interior and the exterior. Our derivation relies
solely on the scalar field equation and the 00--component of the
field equation, without the need for the full set of equations, specifically
the $11-$ and $22-$ components of the field equation. Consequently,
the conservation equation (by way of the Bianchi identity) is not
required. (iii)\emph{$\ $Universality of result:} The final formula,
Eq. \eqref{eq:gamma-exact-1}, holds for all field strengths and all
types of matter (whether convective or non-convective, for example).
We do not assume the matter comprising the stars to be a perfect fluid
or isentropic. The only constraints on matter come from the stationary
and spherical symmetry through Eq. \eqref{eq:EMT}.\vskip12pt

\emph{Higher-derivative characteristics}---In contrast to the one-parameter
Schwarzschild metric, the Brans Class I solution depends on two parameters,
i.e. the solution is not only defined by its gravitational mass, but
also by \emph{a scalar mass} besides the gravitational one \citep{Bronnikov-1973}.
This is because the BD description of gravity involves more fields
than the only metric involved in GR, a scalar field being also part
of the gravitational sector. (The same is also to be expected in the
higher order theories framework, like $f(\mathcal{R})$ theories,
since $f(\mathcal{R})$ theories can be recast as $\omega=0$ BD theories
endowed with a scalar dependent potential.) The exterior BD vacuum
should reflect the internal structure and composition of the star.
This expectation is confirmed in the final result, Eq. \eqref{eq:gamma-exact-1},
which underscores the participation of the parameter $\Theta$.\vskip12pt

\begin{figure}[!t]
\noindent \begin{centering}
\hskip-10pt\includegraphics[scale=0.6]{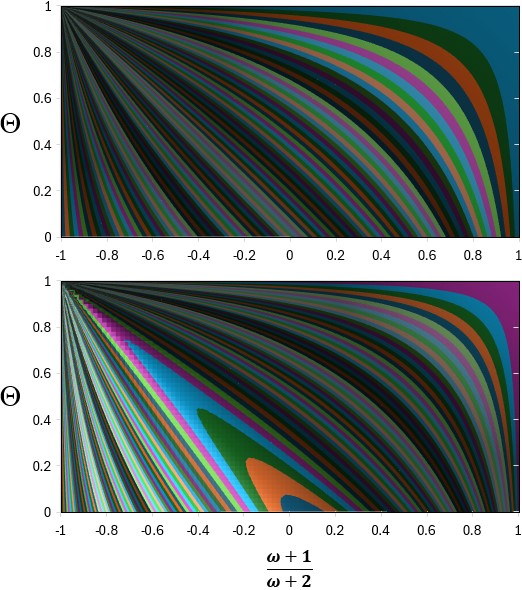}
\par\end{centering}
\caption{\label{fig:gamma-countour}Contour plots of $\gamma_{\,\text{exact}}$
(upper panel) and $\delta_{\,\text{exact}}$ (lower panel) in terms
of $\Theta$ and $\frac{\omega+1}{\omega+2}$, for the range of $\Theta\in[0,1]$
and $\omega\in(-3/2,+\infty)$, the latter corresponding to $\frac{\omega+1}{\omega+2}\in(-1,1)$.
A measured $\gamma_{\,\text{exact}}\approx1$ could mean $\frac{\omega+1}{\omega+2}\approx1$
(i.e., $\omega\gg1$) or $\Theta\simeq1^{-}$ (i.e., ultra-relativistic
matter). Contours are equally spaced in 0.02 increment. For a given
contour in the upper panel, the corresponding value of $\gamma_{\,\text{exact}}$
can be read on the abscissa axis, where the contour intersects it.
By measuring \emph{both} $\gamma$ and $\delta$, the values of $\omega$ and $\Theta$ may be evaluated.}
\end{figure}

\emph{Role of pressure}---Figure \ref{fig:gamma-countour}
shows contour plots of $\gamma_{\,\text{exact}}$ and $\delta_{\,\text{exact}}$
as functions of $\gamma_{\,\text{PPN}}$ (i.e, $\frac{\omega+1}{\omega+2}$)
and $\Theta$. The parameter $\omega$ lies in the range $(-3/2,+\infty)$
(to avoid the ``phantom'' kinetic term for the BD scalar field when
moving to the Einstein frame, an issue mentioned at the beginning
of Section \ref{sec:EMT}) which translates to $\frac{\omega+1}{\omega+2}\in(-1,1)$.
The parameter $\Theta$ lies in the range $[0,1)$ in which $0$ corresponds
to ``Newtonian'' stars, and $1$ corresponds ultra-relativistic matter
where $p_{\parallel}=p_{\perp}=\frac{1}{3}\epsilon$. It is important
to note that the strict upper bound $\Theta=1$ is excluded since
a linear equation of state for matter cannot produce a well-defined
star surface, at which location $p=0$, when solving the Tolman--Oppenheimer--Volkoff
(TOV) equation for GR and BD stars.\vskip4pt

In the ranges for $\omega$ and $\Theta$ given above,
the respective ranges for $\gamma$ and $\delta$ are $\gamma\in(-1,1]$
and $\delta\in[-11/64,4/3)$. The $\delta$ parameter approaches its
upper bound of $4/3$ at $\omega=-3/2$ and $\Theta\rightarrow1^{-}$,
and its lower bound of $-11/16$ at $\{\omega=-14/15,\ \Theta=0\}$.\vskip4pt

There are three interesting observations:
\begin{itemize}
\item An ultra-relativistic limit, $\Theta\simeq1^{-}$, would render $\gamma_{\,\text{exact}}\simeq1$
and $\delta_{\,\text{exact}}\simeq1$, \emph{regardless} of $\omega$.
\item For Newtonian stars, i.e. low pressure ($\Theta\approx0$), the PPN
$\gamma$ result is a good approximation \emph{regardless} of the
field strength.
\item To evaluate $\omega$ and $\Theta$ for a given star
system, measurements of both $\gamma$ and $\delta$ are necessary.
Since the contour plots of $\gamma$ and $\delta$ in Fig. \ref{fig:gamma-countour}
are not align, it is possible in principle to extract $\{\omega,\ \Theta\}$
from $\{\gamma,\ \delta\}$. A slight complexity might arise: due
to the non-linear dependencies of $\{\gamma,\ \delta\}$ on $\{\omega,\ \Theta\}$
via \eqref{eq:gamma-exact-1} and \eqref{eq:delta-exact-1}, for a
given pair of $\{\gamma,\ \delta\},$multiple choices for $\{\omega,\ \Theta\}$
can exist. Resolving the multiplicity would require additional information,
such as the measurement of another PN parameter. See Section \ref{sec:Higher-PPN}.
\end{itemize}
\vskip4pt

\emph{${\cal O}\left(1/\sqrt{\omega}\right)$ anomaly}---It has been
discovered that the Brans-Dicke-Maxwell (electro)vacuum does not necessarily
converge to a vacuum of GR \citep{Faraoni-1998,Faraoni-1999,Romero-1998,Romero-1993-a,Romero-1993-b,Romero-1993-c,Faraoni-2019,Bhadra-2001,Bhadra-2002,Banerjee-1997}.
Additionally, the convergence rate is $1/\sqrt{\omega}$ rather than
the usual $1/\omega$ behavior. Recently, the present authors proved
that this ${\cal O}\left(1/\sqrt{\omega}\right)$ anomaly can exist
even for non vanishing trace of the EMT \citep{Nguyen-2023-BDKG}
(see also \citep{Chauvineau-2003,Chauvineau-2007}). For this anomaly
to occur, the ``remnant'' BD scalar field needs to exhibit a singularity
or time-dependence. However, these conditions are not satisfied for
static stars, where regularity requirements are imposed. Therefore,
the violation of $\gamma$ reported in this current article is not
related to the ${\cal O}\left(1/\sqrt{\omega}\right)$ anomaly.\vskip12pt

\emph{On the loss of Birkhoff's theorem---}It can be argued that,
in BD gravity, the loss of Birkhoff's theorem and the dependence of
$\gamma$ on a star's internal structure may be interconnected. Indeed,
let us consider, in BD gravity, a static spherisymmetric Newtonian
star (initial state). Its exterior vacuum is described by a Brans
Class I solution given by Eq. \eqref{eq:unify}, characterized by
2 parameters $M_{1}$ and $\Lambda$ (or equivalently $\kappa$ per
Eq. \eqref{eq:kappa}). Note that the parameter $\Lambda$ depends
on the pressure content of the star, as is evident in Eq. \eqref{eq:Lambda-formula}.
For the initial Newtonian star, since $p\ll\epsilon$ inside the whole
star, $\Theta_{\,\text{ini}}$ is approximately zero, rendering $\Lambda_{\,\text{ini}}\approx-\frac{1}{\omega+2}$.
Let us now consider that this star starts collapsing, and that the
collapse ends at some (final) compact state. The pressure is no longer
negligible with respect to $\epsilon$ in this final state, in such
a way that the final value $\Theta_{\,\text{fin}}$ is significant,
making $\Lambda_{\,\text{fin}}$ significantly differing from $\Lambda_{\,\text{ini}}$.
On the other hand, Birkhoff's theorem in GR mandates that any spherisymmetric
vacuum must be independent of the coordinate $t$ regardless of the
\emph{(time-) evolution} of the source. If Birkhoff's theorem were
valid for BD gravity, the vacuum exterior to the collapsing star would
have been left unchanged during the process (i.e. a time independent
solution), which is incompatible with the observation that the final
$\Lambda_{\,\text{fin}}$ differs from the initial one $\Lambda_{\,\text{ini}}$.
Thence, the fact that $\Lambda$\emph{ explicitly depends on $\Theta$},
as described by Eq. \eqref{eq:Lambda-formula}, implies that Birkhoff's
theorem cannot hold for BD gravity. Reciprocally, the strong necessity
to revisit BD gravity's PPN $\gamma$ expression could have been anticipated
from the mere fact that Birkhoff's theorem is not valid in BD gravity.\vskip12pt

\emph{Brans--Dicke stars with ultra-relativistic
matter}---Generally speaking, the limit $\Theta\rightarrow1^{-}$
renders $\Lambda\rightarrow0$ and $\kappa\rightarrow1$ (per Eq.
\eqref{eq:kappa}) \emph{regardless} of $\omega$ in the range $(-3/2,+\infty)$. From Eq. \eqref{eq:unify},
the Brans Class I solution \emph{degenerates} to the Schwarzschild solution in isotropic coordinates
\begin{equation}
\left\{ \begin{array}{rl}
ds^{2} & ={\displaystyle -F(\rho)dt^{2}+G(\rho)\left(d\rho^{2}+\rho^{2}d\Omega^{2}\right)}\\
F(\rho) & {\displaystyle =\left(\frac{1-\frac{M_{1}}{2\rho}}{1+\frac{M_{1}}{2\rho}}\right)^{2}}\\
G(\rho) & ={\displaystyle \left(1+\frac{M_{1}}{2\rho}\right)^{4}}\\
\phi(\rho) & =1
\end{array}\right.
\end{equation}
The degeneracy infers that ultra-relativistic Brans-Dicke stars and
GR stars are \emph{indistinguishable},
as far as their exterior vacua are concerned. Furthermore, as mentioned
in the Introduction section, per Hawking's no-hair theorem \citep{Hawking-1972-BD,SotiriouFaraoni-2012},
non-rotating uncharged BD black holes also correspond to the Schwarzschild
solution. Therefore, considering a static spherisymmetric exterior
vacuum, one cannot distinguish among a GR star, an ultra-relativistic
BD star, and a BD black hole.\vskip8pt

The degeneracy of the Brans Class I solution to the
Schwarzschild solution can be explained by the following observation:
For ultra-relativistic matter, the trace of the EMT vanishes, per
Eq. \eqref{eq:EMT-trace}. The scalar equation \eqref{eq:BD-eq-phi}
then simplifies to $\square\,\Phi=0$ \emph{everywhere}.
Coupled with the regularity condition at the star center, this ensures
a constant $\Phi$ throughout the spacetime which is now described
by the Schwarzschild solution. Consequently, the scalar degree of
freedom in BD gravity is suppressed in the ultra-relativistic limit.
This prompts an intriguing possibility whether Birkhoff's theorem
is fully restored in this limit.\vskip4pt

Interestingly, the degeneracy appears to manifest
in a special type of Bergmann--Wagoner theory \citep{Bergmann-1968,Wagoner-1970}.
In this theory, $\omega$ is promoted to be a function of the scalar
field, $\omega(\Phi)$, and a potential term of the scalar field $\Lambda(\Phi)$
is introduced. The scalar equation in this theory is
\begin{equation}
\nabla_{\mu}\left(\sqrt{2\omega(\Phi)+3}\,\nabla^{\mu}\Phi\right)=\frac{8\pi T+2\Phi^{3}\left(\Lambda(\Phi)/\Phi\right)'}{\sqrt{2\omega(\Phi)+3}}
\end{equation}
where the prime denotes derivative with respect to $\Phi$. For the
choice $\Lambda(\Phi)=\lambda_{0}\,\Phi$ with $\lambda_{0}\in\mathbb{R}$,
the scalar field equation then simplifies to $\nabla_{\mu}\left(\sqrt{2\omega(\Phi)+3}\,\nabla^{\mu}\Phi\right)=0$
for ultra-relativistic matter, $T=0$. This equation obviously admits
a constant field $\Phi\equiv\Phi_{0}$ as a solution, thereby reducing
the metric equation to the standard Einstein-Hilbert equation augmented
with a cosmological term $\lambda_{0}\,g_{\mu\nu}$. The case of $\lambda_{0}=0$
corresponds to the \emph{massless} Bergmann--Wagoner theory.

\section{\label{sec:Closing}Closing and Outlooks}

In closing, we have derived the exact analytical formulae, Eqs. \eqref{eq:gamma-exact-1}
and \ref{eq:delta-exact-1}, for the PN parameters $\gamma$ and $\delta$
for spherical compact mass sources in BD gravity. The derivation's
success hinges on the integrability of the $00-$component of the
field equation, rendering it non-perturbative and applicable for any
field strength and type of matter constituting the source. The derivation
is \emph{parsimonious}, requiring only a subset of the Brans-Dicke
field equations, whereas the full set would be needed to determine
the \emph{interior} of the stars. A complete account of the technical
insights was provided in the preceding section.\vskip18pt

\textbf{\emph{Outlook \#1}}---The conventional PPN result for BD
gravity $\gamma_{\,\text{PPN}}=\frac{\omega+1}{\omega+2}$ lacks dependence
on the physical features of the mass source, a trait shared by other
alternative theories beyond GR, such as the Bergman--Wagoner theory.
In the light of our exact result, the $\gamma_{\,\text{PPN}}$ should
be regarded as an approximation for stars in modified gravity under
low-pressure conditions. Our findings underscore the limitations of
the PPN formalism, particularly in scenarios characterized by high
star pressure. It is plausible to expect that the role of pressure
may extend to other modified theories of gravitation, such as in the massive Brans--Dicke gravity considered in \citep{PRD-1,PRD-2}.\vskip18pt

\textbf{\emph{Outlook \#2}}---The energy and pressure integrals,
$E^{*}$ and $P_{\parallel}^{*}+2P_{\perp}^{*}$ defined in Eq. \eqref{eq:integrals-def},
can be evaluated within the TOV framework for BD gravity. This approach
entails developing the TOV equations and integrating them from the
star's center outward. The resulting exterior solution is fully determined
by factors such as the equation of state of the star's constituent
matter and the central pressure of the star. Accordingly, the matching
between the interior spacetime solution and the exterior vacuum is
automatically handled when the integration crosses the surface at
which the pressure vanishes. However, it should be noted that the
integration procedure is numerical in nature, as there is currently
no exact analytical solution available for the interior of stars in
BD gravity. Inspired by our exact $\gamma$ derivation presented in
this paper, details regarding the TOV equation for BD gravity, presented
in a new optimal gauge choice, are currently underway \citep{Nguyen-compact-star-2}.
\begin{acknowledgments}
We thank the anonymous referee for their constructive comments. HKN
thanks Mustapha Azreg-A\"inou for stimulating discussions, and Valerio
Faraoni, Tiberiu Harko, Peter V\'an, L\'aszl\'o Gergely, Viktor Toth, and the
participants of the XII Bolyai--Gauss--Lobachevsky Conference (BGL-2024):
Non-Euclidean Geometry in Modern Physics and Mathematics for valuable
commentaries. BC thanks Antoine Strugarek for helpful correspondences.
\end{acknowledgments}

\begin{center}
-----------------$\infty$-----------------
\par\end{center}

\appendix

\section{\label{sec:Mapping-Brans-solutions}$\ $Mapping the unified Brans
solution to harmonic radial coordinate}

The uniqueness of Brans solutions can also be made more transparent
with the use of harmonic radial coordinates. This Appendix provides
a brief explanation of this aspect.\vskip4pt

Introducing a new parameter $\chi$ and making the following coordinate
transformation to a radial coordinate $u$
\begin{equation}
\rho=\frac{1}{2}M_{1}\sqrt{\kappa}\frac{1+e^{\sqrt{\chi}u}}{1-e^{\sqrt{\chi}u}};\ \ \ \left(\frac{d\rho}{du}\right)^{2}=\frac{M_{1}^{2}\kappa\chi
}{\left(1-e^{\sqrt{\chi}u}\right)^{4}}
\end{equation}
Resulting in
\begin{align}
\left(1-\frac{M_{1}^{2}\kappa}{4\rho^{2}}\right)^{2}\rho^{2} & =\frac{M_{1}^{2}\kappa}{\sinh^{2}(\sqrt{\chi}u)};\\
\left(1-\frac{M_{1}^{2}\kappa}{4\rho^{2}}\right)^{2}\left(\frac{d\rho}{du}\right)^{2} & =\frac{M_{1}^{2}\kappa\,\chi}{\sinh^{4}(\sqrt{\chi}u)}.
\end{align}
Introducing two parameters $\alpha$ and $\beta$ such that 
\begin{align}
\chi & =(a+b)\,\kappa\label{eq:chi-def}\\
\Lambda & =-\frac{2b}{a+b}\label{eq:Lambda-def}
\end{align}
The BD scalar field and the metric become (upon rescaling $dt\rightarrow\frac{M_{1}\sqrt{\kappa}}{\sqrt{\chi}}dt$
and $ds\rightarrow\frac{M_{1}\sqrt{\kappa}}{\sqrt{\chi}}ds$), respectively
\begin{align}
\phi(u) & =e^{-2bu}\\
ds^{2} & =\frac{1}{\phi(u)}\biggl[-e^{2au}dt^{2}+e^{-2au}\left(\frac{du^{2}}{s^{4}(\chi,u)}+\frac{d\Omega^{2}}{s^{2}(\chi,u)}\right)\biggr]
\end{align}
with the ``Bronnikov'' function (first introduced in \citep{Bronnikov-1973})
defined as
\begin{equation}
s(\chi,\,u):=\frac{1}{\sqrt{\chi}}\sinh\sqrt{\chi}u
\end{equation}
The proper part satisfies the radial harmonic gauge. Depending on
whether $\chi$ is positive, null, or negative, the function is \emph{numerically}
equal to 
\begin{equation}
s(\chi,\,u)=\begin{cases}
\ \frac{1}{\sqrt{\chi}}\sinh\sqrt{\chi}u & \text{if }\chi>0\\
\ u & \text{if }\chi=0\\
\ \frac{1}{\sqrt{-\chi}}\sin\sqrt{-\chi}u & \text{if }\chi<0
\end{cases}\label{eq:my-s}
\end{equation}
The three cases correspond to Brans Class I, Class IV, and Class II,
in that order. Plugging Eqs. \eqref{eq:chi-def} and \eqref{eq:Lambda-def}
in to Eq. \eqref{eq:kappa}, we obtain the following ``constraint''
among the parameters
\begin{equation}
\chi=a^{2}+(2\omega+3)\,b^{2}
\end{equation}
Obviously, a pair of $\{\omega,\,a/b\}$ uniquely specifies the signum
for $\chi.$ This is nothing but the metric obtained in the harmonic
radial coordinate $u$, presented in Bronnikov \citep{Bronnikov-1973}.
The proper part of the metric is the Fisher-Janis-Newman-Winicour
(FJNW) solution \citep{Faraoni-2018,Wyman-1981,Bergmann-1957,Buchdahl-BD-1972,Fisher-1948,Janis-1968}.)
The generality of this metric is thus established therein. For $\omega>-3/2$,
$\chi>0$ is the Brans Class I. Whereas $\omega<-3/2$, all three
possibilities for $\chi$ positive, null, negative are admissible.
However, for a given pair of $\{\omega,\,a/b\}$, the signum of $\chi$
is uniquely determined, selecting the class that the solution belongs.
The ambiguity in picking one solution among the three possibilities
is removed.\vskip4pt

One final remark is that Bronnikov used a slightly different notation
which may have obscured the ``uniqueness'' of the Brans solutions
\citep{Bronnikov-1973})
\begin{align}
ds^{2} & =\frac{1}{\phi(u)}\biggl[-e^{2au}dt^{2}\nonumber \\
 & +e^{-2au}\biggl(\frac{du^{2}}{s_{\text{Bronnikov}}^{4}(k,u)}+\frac{d\Omega^{2}}{s_{\text{Bronnikov}}^{2}(k,u)}\biggr)\biggr]
\end{align}
depending on the signum of $k$:
\begin{equation}
s_{\,\text{Bronnikov}}(k,\,u):=\begin{cases}
\ k^{-1}\sinh ku & \text{if }k>0\\
\ u & \text{if }k=0\\
\ k^{-1}\sin ku & \text{if }k<0
\end{cases}\label{eq:Bronnikov-s}
\end{equation}
The three cases correspond to Brans Class I, Class IV, and Class II,
in that order.\vskip8pt

The distinction between \eqref{eq:my-s} and \eqref{eq:Bronnikov-s}
is subtle. Comparing the two functions, $k$ can be formally identified
with $\sqrt{-\chi}$. However, with $\chi\in\mathbb{R}$, $\sqrt{-\chi}$
can be real or pure imaginary, whereas the $k$ parameter in \eqref{eq:Bronnikov-s}
needs be defined separately for $\chi>0$ and $\chi<0$. The ``bifurcation''
at $k=0$ may have obscured the duality between Brans Class I and
Class II, an issue we elucidated in the Section \ref{sec:Unique}.

\end{document}